\documentclass[lefttitle,3p,review]{elsarticle}

\usepackage{siunitx}
\usepackage[super]{nth}
\usepackage{subcaption}
\usepackage{placeins}
\usepackage[]{natbib}

\usepackage{todonotes}
\usepackage{soul}

\begin{document}

    \newcommand\chips{\textsc{Chips}}
    \DeclareSIUnit{\littleLitre}{l}

    \begin{frontmatter}
    
        \title{The Design and Construction of the \textsc{Chips} Water Cherenkov Neutrino Detector}

        \author[2] {B.~Alonso~Rancurel}
        
        \author[5] {N.~Angelides}
        
        \author[2] {G.~Augustoni}
        
        \author[1] {S.~Bash\corref{cor1}\footnote{Now at: Technical University of Munich, James-Franck-Straße, 85748, Garching, Germany}}
        \ead{simeon.bash@ucl.ac.uk}
        \cortext[cor1]{Corresponding author}
        
        \author[3] {B.~Bergmann}
        
        \author[4] {N.~Bertschinger}
        
        \author[1] {P.~Bizouard}
        
        \author[1] {M.~Campbell}
        
        \author[19] {S.~Cao}
        
        \author[4] {T.~J.~Carroll}
        
        \author[1] {R.~Castellan}
        
        \author[14] {E.~Catano-Mur}
        
        \author[12] {J.~P.~Cesar}
        
        \author[22] {J.~A.~B.~Coelho}
        
        \author[28] {P.~Dills}
        
        \author[1] {T.~Dodwell}  
        
        \author[1] {J.~Edmondson}
        
        \author[7] {D.~van~Eijk}
        
        \author[4] {Q.~Fetterly}
        
        \author[2] {Z.~Garbal}
        
        \author[1,23] {S.~Germani}
        
        \author[1] {T.~Gilpin}
        
        \author[2] {A.~Giraudo}
        
        \author[11] {A.~Habig}
        
        \author[4] {D.~Hanuska}
        
        \author[9] {H.~Hausner}
        
        \author[4] {W.~Y.~Hernandez}
        
        \author[20] {A.~Holin}
        
        \author[18] {J.~Huang}
        
        \author[1] {S.~B.~Jones}
        
        \author[16] {A.~Karle}
        
        \author[1] {G.~Kileff}
        
        \author[1] {K.~R.~Jenkins}
        
        \author[7] {P.~Kooijman}
        
        \author[9] {A.~Kreymer}
        
        \author[4] {D.~A.~Loving} 
        
        \author[4] {G.~M.~LaFond}
        
        \author[12] {K.~Lang}
        
        \author[16] {J.~P.~Lazar}
        
        \author[10] {R.~Li}
        
        \author[24,25] {K.~Liu}
        
        \author[1,3] {P.~Mánek}
        
        \author[13] {M.~L.~Marshak}
        
        \author[13] {J.~R.~Meier}
        
        \author[13] {W.~Miller}
        
        \author[14]{J.~K. Nelson}
        
        \author[27]{C.~Ng}
        
        \author[1]{R.~J.~Nichol}
        
        \author[21] {V.~Paolone}
        
        \author[1] {A.~Perch}
        
        \author[1] {M.~M.~Pfützner}
        
        \author[1,14] {A.~Radovic} 
        
        \author[8]{K.~Rawlins}
        
        \author[4] {P.~Roedl}
        
        \author[4] {L.~Rogers}
        
        \author[15]{I.~Safa}
        
        \author[17]{A.~Sousa}
        
        \author[1] {J.~Tingey}
        
        \author[1] {J.~Thomas}
        
        \author[14] {J.~Trokan-Tenorio}
        
        \author[14] {P.~Vahle}
        
        \author[6] {R.~Wade}
        
        \author[4] {C.~Wendt}
        
        \author[26] {D.~Wendt}
        
        \author[1]{L.~H.~Whitehead 
        \footnote{Now at: Cavendish Laboratory, University of Cambridge, Cambridge CB3 0HE, United Kingdom}}
        
        \author[4] {S.~Wolcott}
        
        \author[16]{T.~Yuan}
        
        \address[1]{Department of Physics and Astronomy, University College London, Gower Street, London, WC1E 6BT, United Kingdom}
        
        \address[2]{Aix-Marseille University, Science Faculty in Saint-Jérôme campus, 13013 Marseille, France}
        
        \address[3]{Institute of Experimental and Applied Physics, Czech Technical University in Prague, Husova 240/5, 110 00  Prague 1, Czech Republic.}
        
        \address[4]{Department of Physics, University of Wisconsin, Madison, WI 53706, USA}
        
        \address[5]{Imperial College London, Physics Department, Blackett Laboratory, London SW7 2AZ, UK}
        
        \address[6]{Avenir Consulting, Abingdon, Oxfordshire, UK}
        
        \address[7]{Nikhef, Science Park 105, 1098 XG, Amsterdam, The Netherlands}
        
        \address[8]{University of Alaska Anchorage, 3211 Providence Dr. Anchorage, AK 99508}
        
        \address[9]{Fermi National Accelerator Laboratory, Batavia, IL 60510, USA}
        
        \address[10]{School of Physics and Astronomy, Shanghai Jiao Tong University, MOE Key Laboratory for Particle Astrophysics and Cosmology, Shanghai Key Laboratory for Particle Physics and Cosmology, Shanghai 200240, China}
        
        \address[11]{Department of Physics and Astronomy, University of Minnesota Duluth, Duluth, Minnesota 55812, USA}
        
        \address[12]{Department of Physics, University of Texas at Austin, Austin, TX 78712, USA}
        
        \address[13]{University of Minnesota, Minneapolis, Minnesota 55455, USA}
        
        % School of Physics and Astronomy, University of Minnesota Twin Cities, Minneapolis, Minnesota 55455, USA
        
        \address[14]{Department of Physics, William \& Mary, Williamsburg, VA 23187, USA}
        
        \address[15]{Department of Physics, Columbia University, New York, NY, USA}
        
        \address[16]{Department of Physics and Wisconsin IceCube Particle Astrophysics Center, University of Wisconsin–Madison, Madison, WI 53706, USA}
        
        \address[17]{Department of Physics, University of Cincinnati, Cincinnati, OH 45221, USA}
        
        \address[18]{School of Physics and Astronomy, Shanghai Jiao Tong University, China}
        
        \address[19]{Institute For Interdisciplinary Research in Science and Education (), ICISE, Quy Nhon, Vietnam}
        
        \address[20]{Particle Physics Department,
        STFC Rutherford Appleton Laboratory, Harwell Campus,  Didcot OX11 0QX, United Kingdom}
        
        \address[21]{University of Pittsburgh, Pittsburgh, PA, 15260, USA}
        
        \address[22]{Université Paris Cité, Astroparticule et Cosmologie, F-75013 Paris, France}
        
        \address[23]{Dipartimento di Fisica e Geologia, Università degli Studi di Perugia, I-06123 Perugia, Italy}
        
        \address[24]{Department of Physics, Sun Yat-sen University, 135 Xingang Xi Road, Haizhu District, Guangzhou, 510275}
        
        \address[25]{Department of Physics, Tsinghua University, 30 Shuangqing Road, Haidian District, Beijing, 100084}
        
        \address[26]{Department of Physics, University of Colorado, Boulder, Colorado 80309, USA}
        
        \address[27]{Department of Physics and Astronomy, Michigan State University, East Lansing, MI 48824, USA}
        
        \address[28]{Department of Mechanical Engineering, University of Wisconsin, Madison, WI 53706, USA}

        \begin{abstract}
            \textsc{Chips} (CHerenkov detectors In mine PitS) was a prototype large-scale water Cherenkov detector located in northern Minnesota. The main aim of the R\&D project was to demonstrate that construction costs of neutrino oscillation detectors could be reduced by at least an order of magnitude compared to other equivalent experiments. This article presents design features of the \textsc{Chips} detector along with details of the implementation and deployment of the prototype. While issues during and after the deployment of the detector prevented data taking, a number of key concepts and designs were successfully demonstrated.
        \end{abstract}
    
    \end{frontmatter}
    
    \newpage
    \tableofcontents
    \section{Introduction}
    \label{sec:introduction}
    
    Neutrino detectors are typically very large and therefore very expensive. Planned and future detectors will need to have masses of tens or even hundreds of kilotons (\SI{}{\kilo\tonne}) to register significant event counts. 
    
    Water Cherenkov detectors, owing to their large footprint and common target volume, can be particularly affected by cosmic backgrounds which necessitates additional shielding. This is commonly resolved by constructing such detectors deep underground, adding significantly to the costs. As an example, the Hyper-Kamiokande project is projected to cost over \$600m for a \SI{258}{\kilo\tonne} detector or \$2.3m per \SI{}{\kilo\tonne} \cite {Castelvecchi2019JapanDetector,DiLodovico2017TheExperiment}. 
    
    The purpose of the \textsc{Chips} (Cherenkov Detectors In Mine Pits) project was to demonstrate, through the construction of a prototype detector, that the cost per \SI{}{\kilo\tonne} for future water Cherenkov (\textsc{wc}) experiments could be lowered by an order of magnitude. To achieve this a \SI{5}{\kilo\tonne} \textsc{Chips} detector would be deployed in a flooded mine pit, \SI{707}{\kilo\metre} away and \SI{7}{\milli\radian} off-axis from the \textsc{N}u\textsc{mi} beam at Fermilab. Crucially the water in the pit would provide both the cosmic ray shielding and the mechanical support for the purified detector water volume, thus avoiding the need for expensive excavation and support structures. The use of commercial, off the shelf components, for water purification, planes of photomultiplier tubes (\textsc{pmt}s) and readout electronics, would further reduce costs.  
    In this paper we describe the main characteristics of the novel detector design along with selected details of their implementation.
    
    \newpage
    \section{Experimental Concept}
    \label{sec:experimentalConcept}
    
    % \begin{figure}
    %     \centering
    %     \includegraphics[width=0.5\textwidth]{final_cosmic_outputs.pdf}
    %     \caption{Components of the bottom end cap. The gusset plates which were used to join pieces together are visible as are the tyres which the frame sits on. The hollow lattice style design saves weight as less steel is required than for a solid detector.}
    %     \label{fig:finalCosmicOutputs}
    % \end{figure}
    
    The \textsc{Chips} concept uses a volume of purified water as the detector medium. The volume is enclosed in a light- and water- tight membrane, submerged in a much larger volume of water such as a lake or mine pit.
    
    Compared to conventional \textsc{wc} detector designs, this architecture offers three key advantages. Firstly, since suitable bodies of water are more likely to occur naturally than deep caverns, expensive excavation is not required. Secondly, as the entire detector is submerged, the water outside supports the water volume contained inside. This means that support structures are only needed to maintain the detector’s shape rather than support the kilotons of its internal mass. Thirdly, the detector could be largely constructed in its final form above ground and later deployed to its submerged location in one piece. The key drawback of this concept is that following deployment, the entire detector is under water including its instrumentation, rather than just the photo-sensitive parts of its sensors. This increases the difficulty of, and sometimes effectively prohibits post-deployment repairs or maintenance.
    
    Despite the fact that \textsc{wc} neutrino detectors are usually constructed deep underground,  considerable effort is still expended in data analysis, classifying events and eliminating cosmic rays from the signal. However, if the detector is focused only on neutrinos coming from an accelerator, significant background reduction is possible even for modest overburdens \cite{Ayres2007}. The \textsc{N}u\textsc{mi} neutrino beam at Fermilab produces neutrinos in \SI{10}{\micro\second}~beam spills. It has been shown that using modern machine learning reconstruction techniques to identify the beam neutrinos, as little as \SI{50}{\metre} of water overburden would be sufficient to discriminate between cosmic ray and neutrino beam events \cite{Tingey2023}. 
    
\begin{figure}
        \centering
        \includegraphics[width=0.75\textwidth]{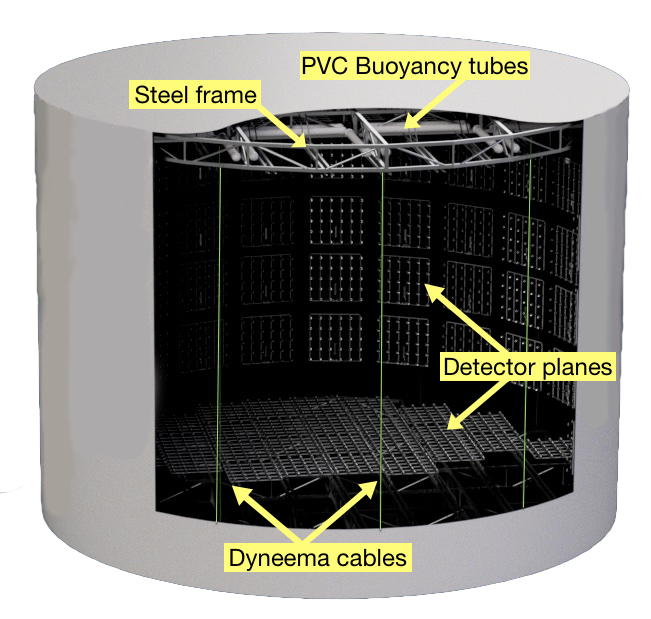}
        \caption{A rendering of the fully deployed \chips{} detector concept showing the overall structure: two steel end caps surrounding a cylindrical volume. The top cap is buoyant because of  the \textsc{pvc}~buoyancy tubes, whereas the bottom cap sinks, exerting tension on the vertical cables. The light-tight liner encloses the detector, protecting instrumented planes from undesirable incoming light. In addition to horizontal planes installed at the time of deployment, the rendering also shows vertical planes, which were envisioned to be installed during a future upgrade along the curved interior wall.}
        \label{fig:chipsRenderLabelled}
    \end{figure}

    To further minimise construction, assembly and maintenance costs, the prototype detector used as many standardised off-the-shelf components and readily available building materials as possible. It was designed in such a way that non-specialists (including students) could perform the majority of the construction tasks safely. An overview of the detector concept can be seen in Figure~\ref{fig:chipsRenderLabelled}.
    
    \section{The PolyMet Mine Site}
    \label{chap:TheCHIPSDetector:sec:ThePolyMetMineSite}
    
\begin{figure}
        \centering 
         \begin{subfigure}[b]{0.49\textwidth}
             \centering
             \includegraphics[width=\textwidth]{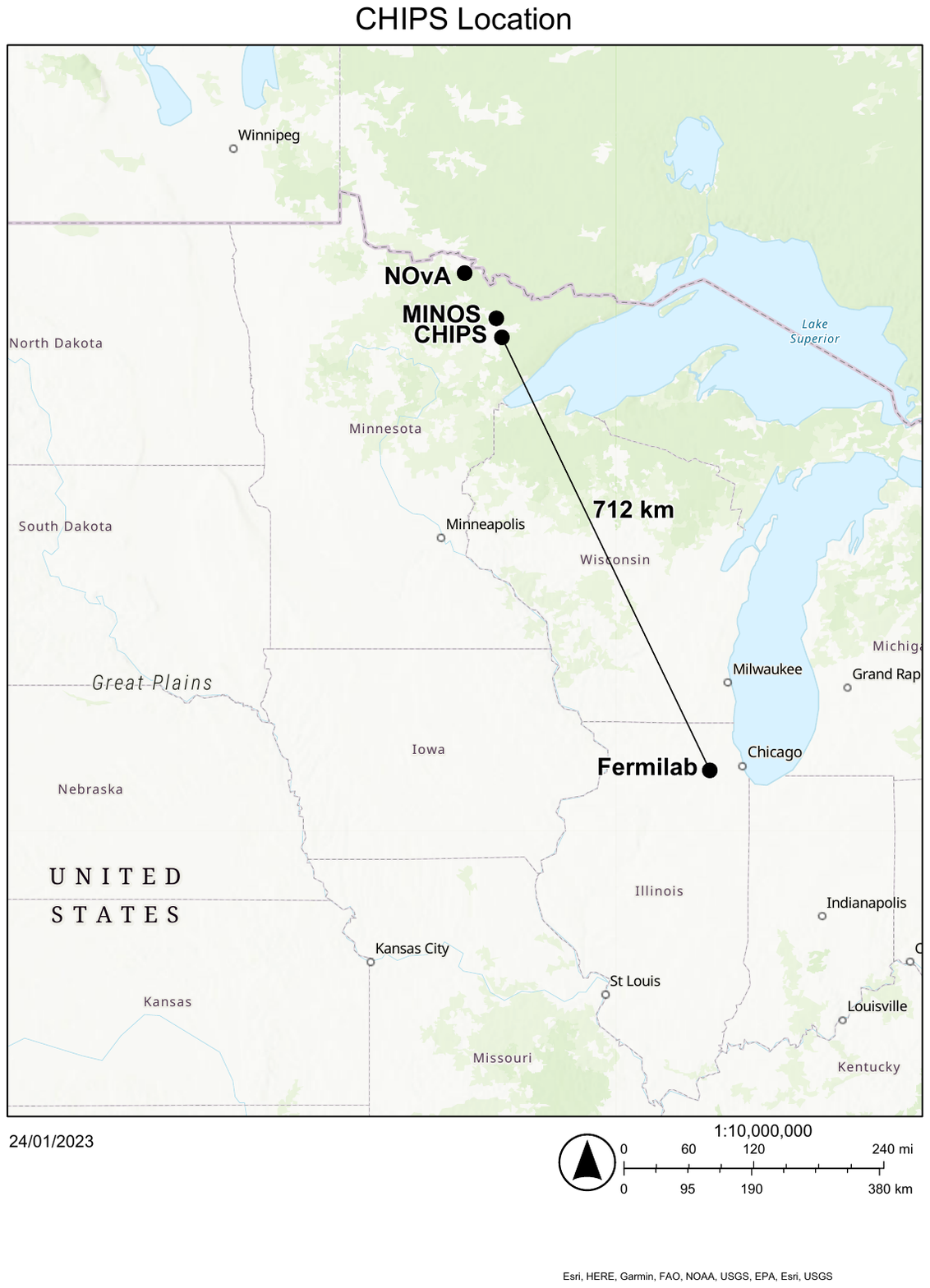}
             \caption{Map showing the locations of the \textsc{N}u\textsc{mi} beam source at Fermilab, the \textsc{w2w} mine pit, where \chips{} was located, and other neutrino detectors (MINOS and NOvA). \cite{Esri2023}.}
             \label{fig:map}
         \end{subfigure}
         \hfill
         \begin{subfigure}[b]{0.49\textwidth}
             \centering        \raisebox{\dimexpr.5\height-.5\baselineskip}{\includegraphics[width=\textwidth]{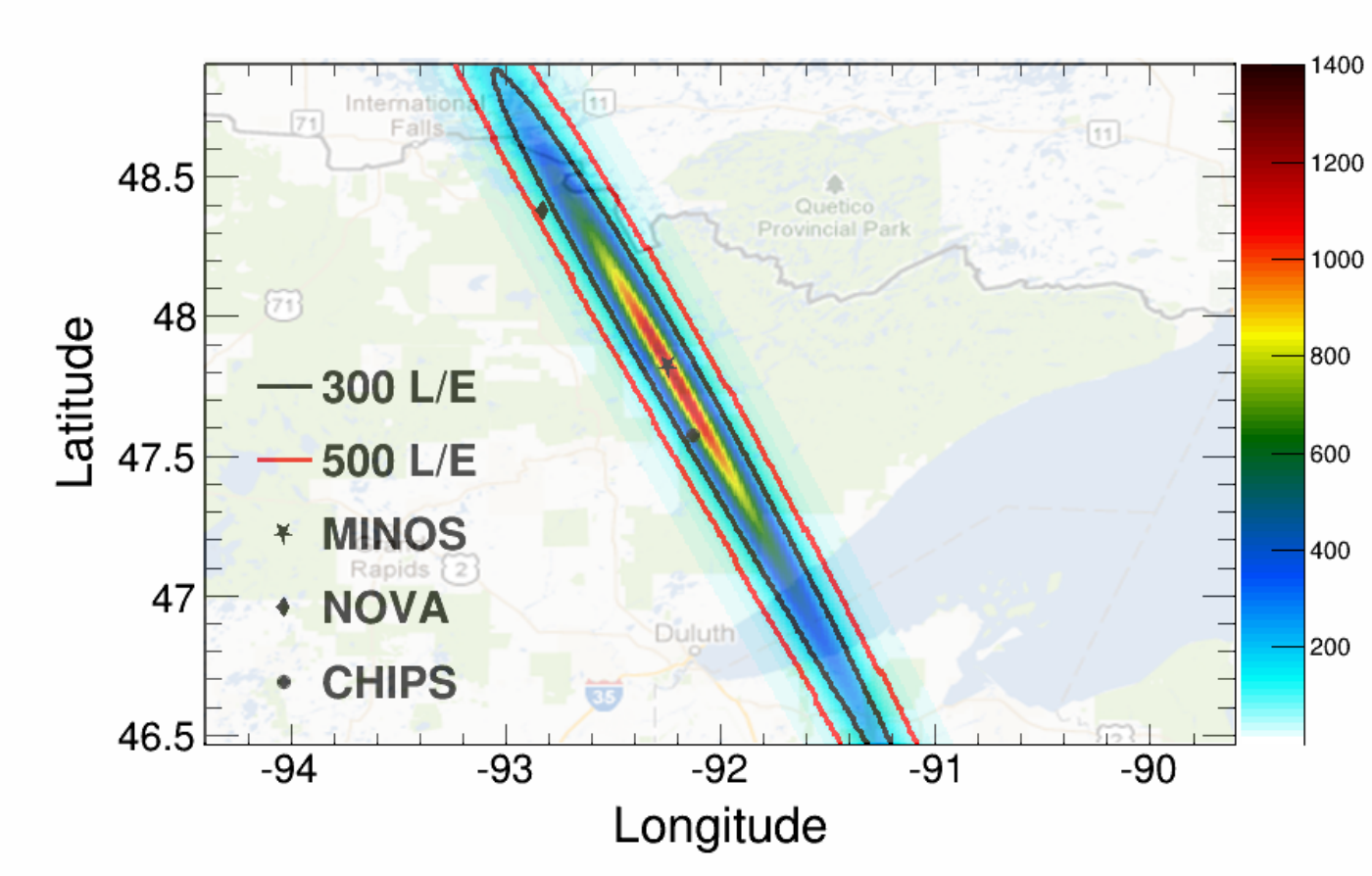}}
             \caption{A map of the \textsc{N}u\textsc{mi} beam exiting the Earth's surface showing the positions of \chips{}, \textsc{No}v\textsc{a} and \textsc{Minos}. The color axis shows the expected number of neutrino events per year per kiloton of water assuming no neutrino oscillations. Contours of constant $L/E$ are shown. Image taken from \cite{Adamson2013CHerenkovFNAL}.}
             \label{fig:NuMI_Map}
        \end{subfigure}
        \vspace {5cm}
        \caption{}
        \label{fig:hadron_prod_uncert}
    \end{figure}

     The \textsc{Chips} prototype was located at Wentworth 2W (\textsc{w2w}), a disused iron quarry at the PolyMet mine site near Hoyt Lakes, Minnesota (shown in Figure \ref{fig:map}). At the time of construction, this pit was still used as a buffer for managing water levels across the larger PolyMet site. This meant that water was occasionally pumped between the \textsc{w2w} pit and other nearby mine pits to moderate their water levels and chemical composition.
    
    The PolyMet site was uniquely useful to \chips{} for several reasons. Firstly, the abandoned and flooded \textsc{w2w} pit provided a suitably deep location (\SI{60}{\metre}) for the detector. Secondly, the existing infrastructure that was maintained for future mining operations included road and rail connections, despite the remoteness of the location. This permitted easy access for heavy machinery and construction industries. The location of the mine site itself was important; the site was within an accessible distance to the Soudan Underground Laboratory Surface Building, which supplied large equipment and tools. 
    Finally, and most importantly, the site was in the path of the \textsc{N}u\textsc{mi} beam as it re-emerged from the Earth (shown in Figure~\ref{fig:NuMI_Map}). 
    
\begin{figure}
        \centering
        \includegraphics[width=0.5\textwidth]{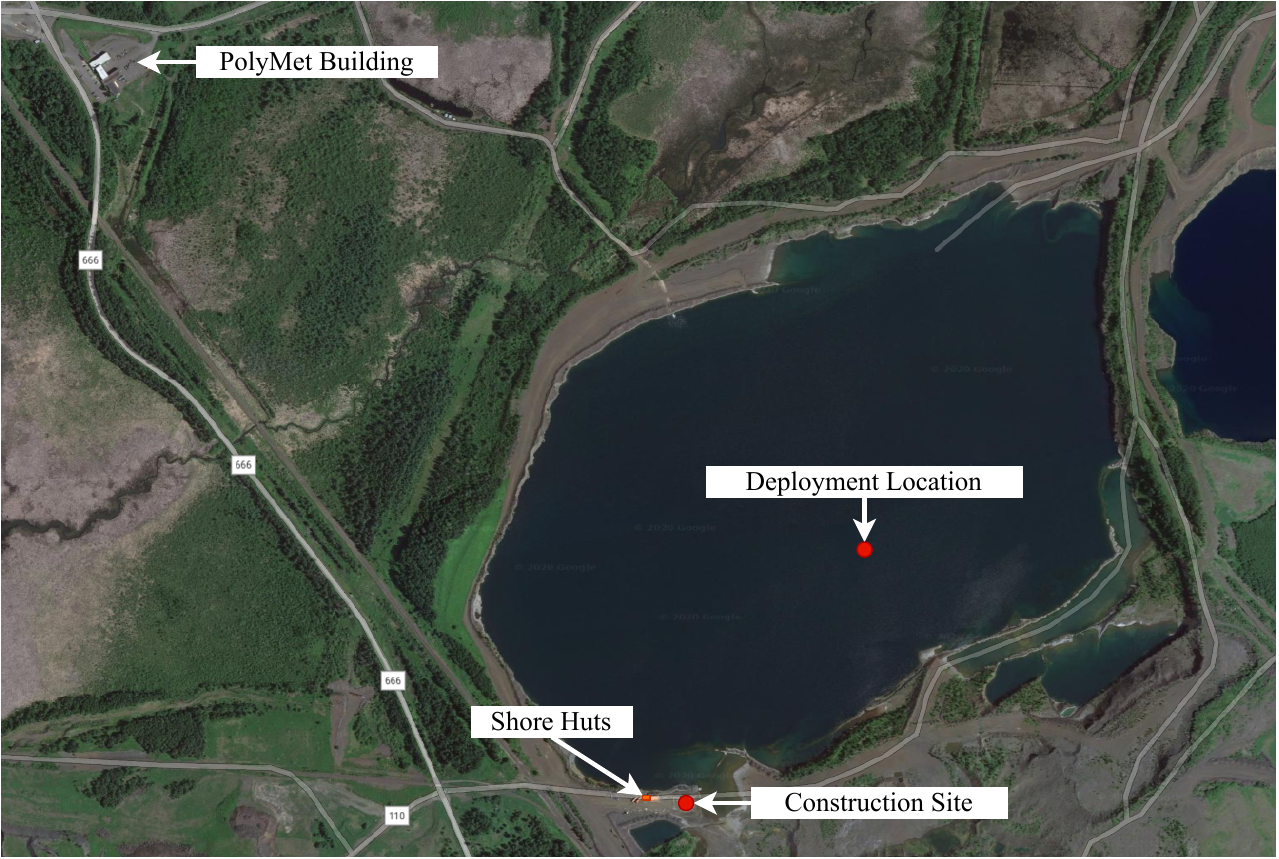}
        \caption{The PolyMet mine site. \chips{} was built beside the mine pit on a flat construction area with additional assembly done in the PolyMet building.  %\cite{Tingey2021ConvolutionalProject}.
        }
        \label{fig:PolyMet_Map}
    \end{figure}

    %The p\textsc{h} level and balance of chemicals in the water of the mine pits were tightly regulated for environmental reasons. As rain water ran into the mine pits through different sections of the mine site, the p\textsc{h} and chemical balance of the water changed. Water was pumped between the different mine pits to maintain safe levels and pH. In autumn, water was pumped into the \chips{} pit and then redistributed over winter months. Therefore, the water level of \textsc{w2w} changed seasonally. 
    The construction of \chips{} exploited the changes of the mine pit's controlled water levels. This allowed the detector to be built on dry land, which was later flooded prior to deployment (this is described in detail in Section~\ref{chap:TheCHIPSDetector:sec:DeploymentProcedure}). The detector and its elements were constructed on a slipway at the edge of the mine pit as shown in Figure~\ref{fig:PolyMet_Map}. Adjacent to this location, two huts were erected to house a water treatment plant and electrical equipment. Space inside the PolyMet main building was dedicated to preparation of computing equipment and light-sensitive instrumentation.
    
    \section{Mechanical structure}
    \label{chap:TheCHIPSDetector:sec:mechanicalStructure}
    
\begin{figure}
        \centering
        \includegraphics[width=\textwidth]{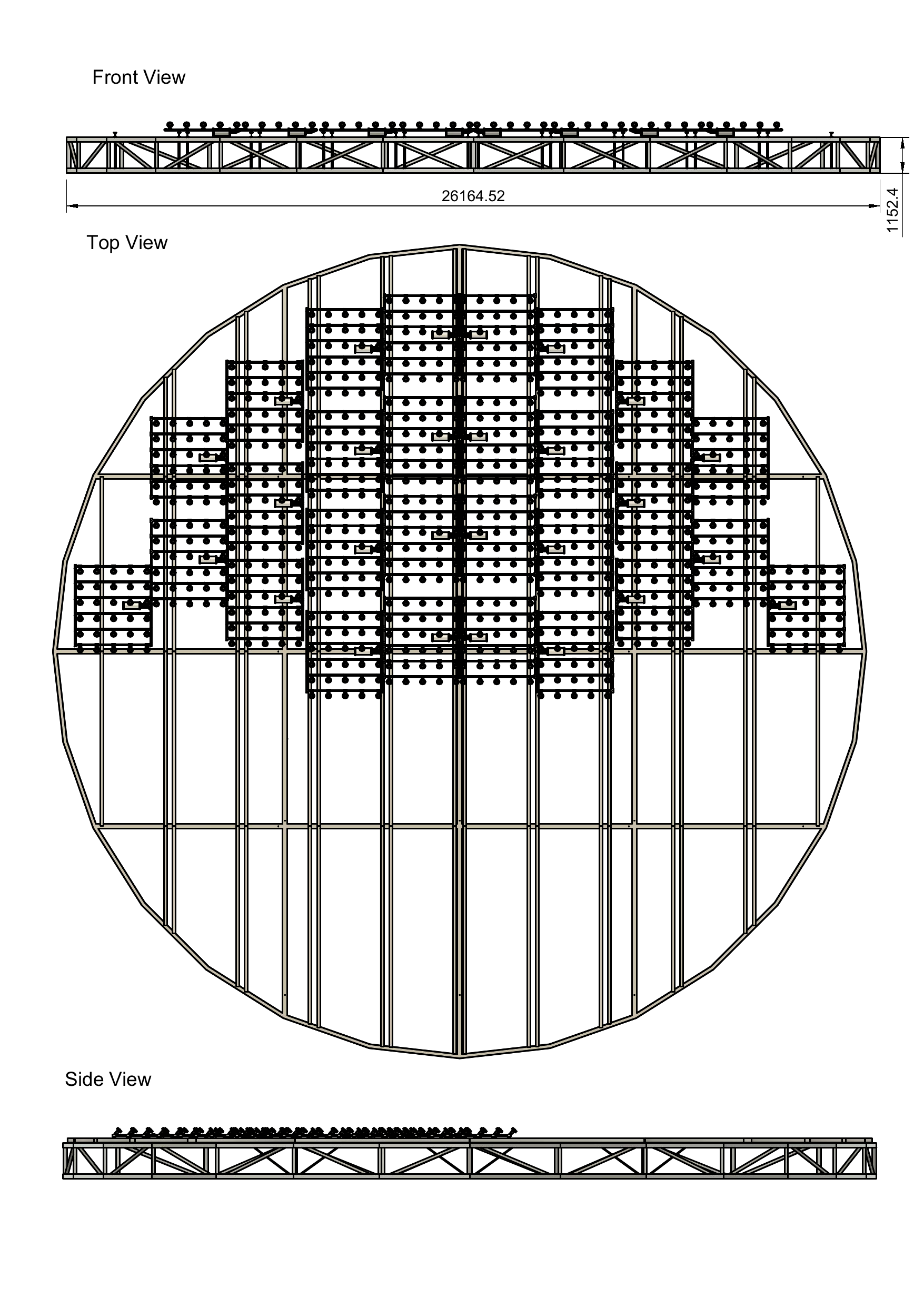}
        \caption{}
        \label{fig:cap_diagram}
    \end{figure}

    The \chips{} detector was mechanically supported by a metal frame comprising two (\SI{25}{\metre}) diameter circular end caps. A technical drawing of an end cap can be seen in figure \ref{fig:cap_diagram}. To achieve a high strength-to-weight ratio, both caps relied on lattice frame construction. As the structure was expected to be submerged in water for extended periods of time, stainless steel was selected because its corrosion resistance would prevent rust from degrading water clarity in the internal volume. Rather than manufacturing perfect circles, the end caps were rigid icosikaioctagon (28-sided regular polygons) steel frames. Steel beams, known as \emph{stringers}, were mounted facing the interior of the cylinder in parallel rows, to which the instrumentation would be attached. Overall each steel end cap weighed \SI{14}{\tonne}.
    
    Before deployment, the frame was fabricated off-site and shipped in pieces to the mine pit. Each piece was assembled with a single telescopic forklift. Stringers could be moved by two workers. This meant that, including the forklift operator, the frame could be assembled by a team with as few as three people. 
    
\begin{figure}
        \centering
        \includegraphics[width=1.0\textwidth]{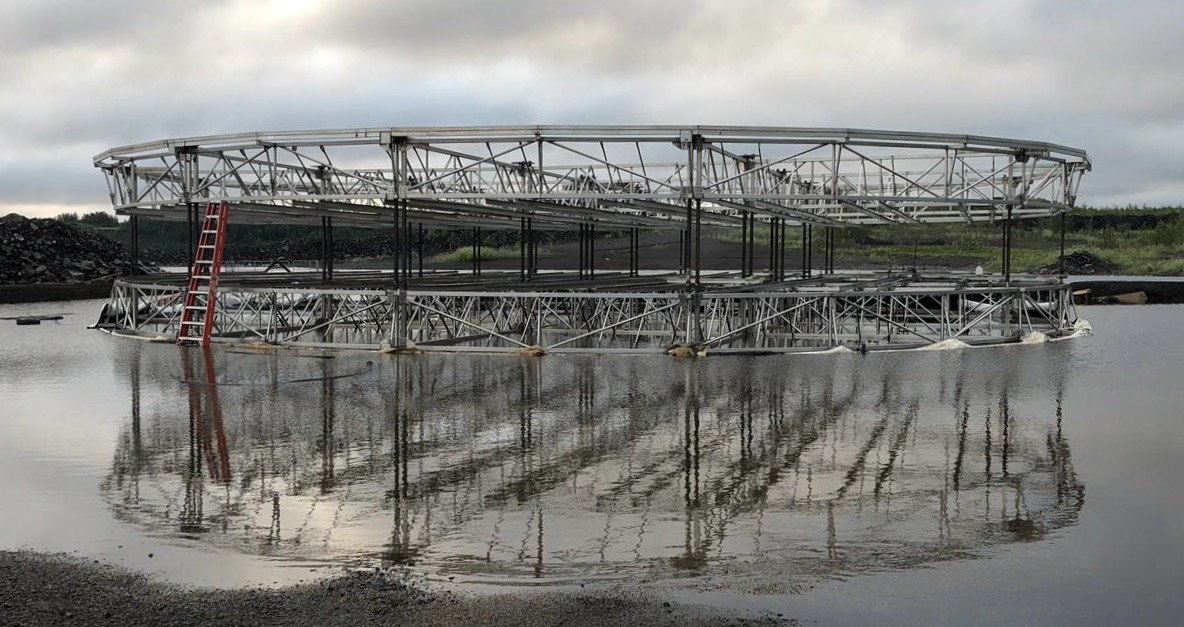}
        \caption{Completed frame. The pillars supporting the top cap are clearly visible.}
        \label{fig:Frame_Completed}
    \end{figure}

\begin{figure}
        \centering
        \includegraphics[width=1.0\textwidth]{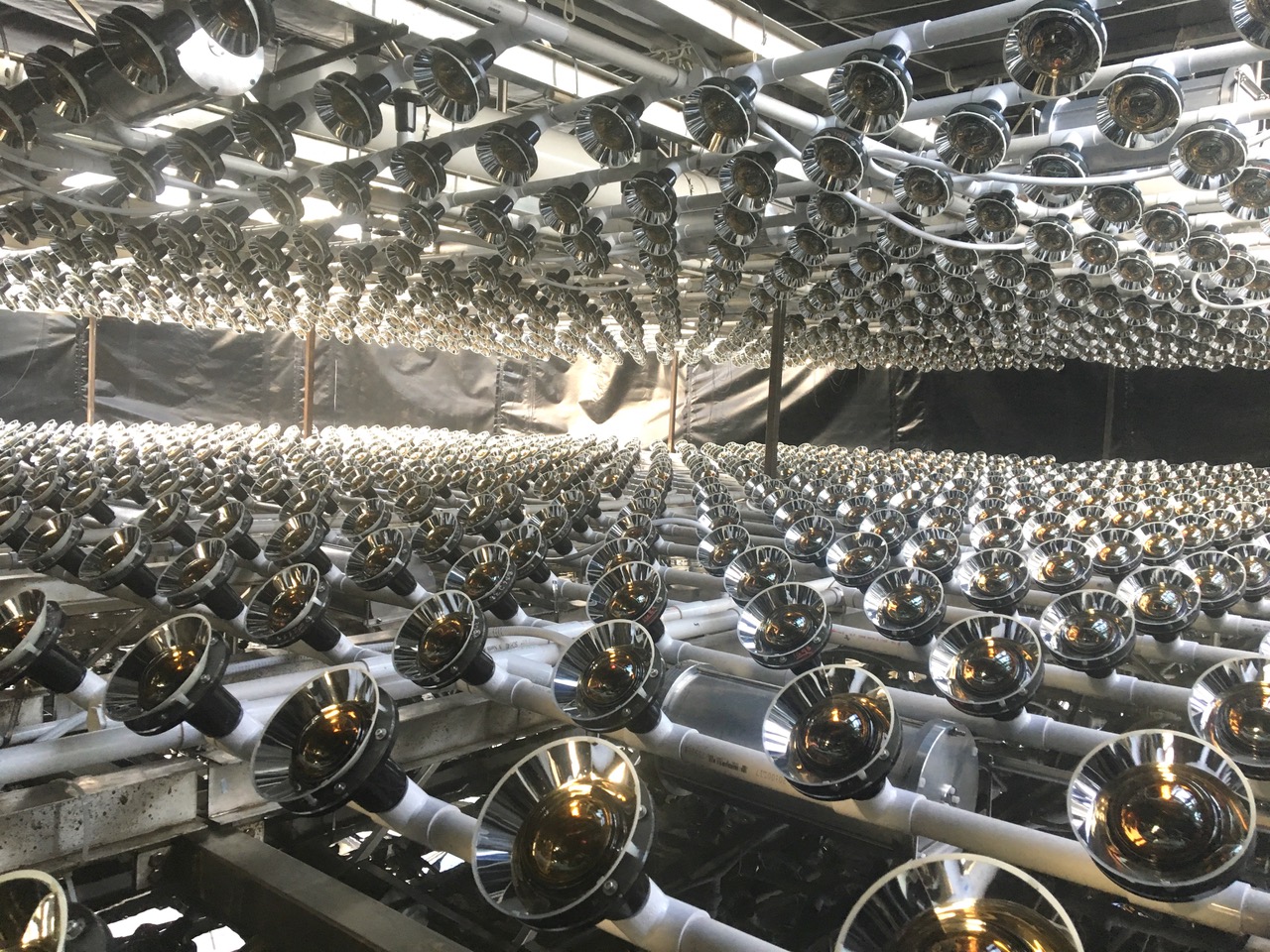}
        \caption{The interior of the detector: the rows of Nikhef \textsc{pmt}s angled to point into the neutrino beam can be seen. The narrow height between the bottom and top caps before the expansion during deployment is visible. Image courtesy of Albrecht Karle.}
        \label{fig:detectorInsideLight}
    \end{figure}

    Both end caps were constructed on top of each other. First, the bottom cap was built on the ground supported on rubber tyres. The bottom cap had 92 stainless steel pegs distributed approximately uniformly across the detector pointing upwards; hollow stilts made of steel tubing were mounted on top of these pegs visible in Figure~\ref{fig:Frame_Completed}. The top cap was built sitting on top of the stilts. During construction, \textsc{pvc} tubes were installed to provide approximately \SI{15}{\tonne} of flotation which made the top cap net buoyant in water as shown in the top of Figure~\ref{fig:detectorInsideLight}.
    
    The completed frame is displayed in Figure~\ref{fig:Frame_Completed}. Following deployment, the two caps were no longer rigidly attached to each other, since the bottom cap sank while the top cap floated. 
    
    In order to accurately measure relatively faint Cherenkov light, the internal detector volume needed to be clear and well-isolated from external light sources. To this end, the detector was fully enclosed in a geomembrane liner. In addition to providing a light seal, this barrier also implemented the separation between internal, filtered water and external pit water, which contained impurities and wildlife. In addition, during deployment the liner would act as a hull that allowed the detector to remain dry inside, float due to water displacement and move like a boat (see Section~\ref{chap:TheCHIPSDetector:sec:DeploymentProcedure} for details).
    
    The liner was crafted from flexible fibre-reinforced plastic (XR-5 Geomembrane \cite{Seaman2023Geomembrane}), conventionally used for roofing. This material provided sufficient strength as well as light- and water-tight properties. The membrane was delivered in \SI{2.54}{\metre}~wide, \SI{183}{\metre}~long rolls. Following construction of the supporting frame, the liner was installed in individual strips, which were heat-fused using specialist welding equipment to form the skin of a \SI{25}{\metre}~diameter, \SI{12.5}{\metre}~tall closed cylinder. Since the height of the detector was expected to increase during deployment, the liner was pooled at the base of the detector. Although the liner was made of a robust material, care was taken to protect it to ensure it remained light tight and waterproof - the entire detector was raised on rubber tyres so the liner wasn't pressed into uneven ground and plastic covers were placed on edges of the steel frame so the liner could not be cut by blowing against the frame in the wind. Finally, liner welds which involved complicated curved geometries or more than two pieces of liner joining at the same place had additional patches welded on top of the joints to ensure quality seals. 
        
    \section{Structure summary and deployment procedure}
    \label{chap:TheCHIPSDetector:sec:DeploymentProcedure}
    
    The steel frame, the liner, the water and the flotation all worked together to give the \textsc{Chips} detector its final structural form. The frame gave the cylinder its rigid round cross-sectional shape and acted as a mounting platform for detector planes. The height of the detector was determined by both the liner and the Dyneema cables, which limited the largest possible distance between end caps. Since the detector was filled with water and was itself submerged underwater, the water on the outside of the detector supported the weight of the water inside its volume. This meant the frame only needed to bear its own weight and the weight of the instrumentation attached to it.
    
\begin{figure}
        \centering
        \includegraphics[width=0.8\textwidth]{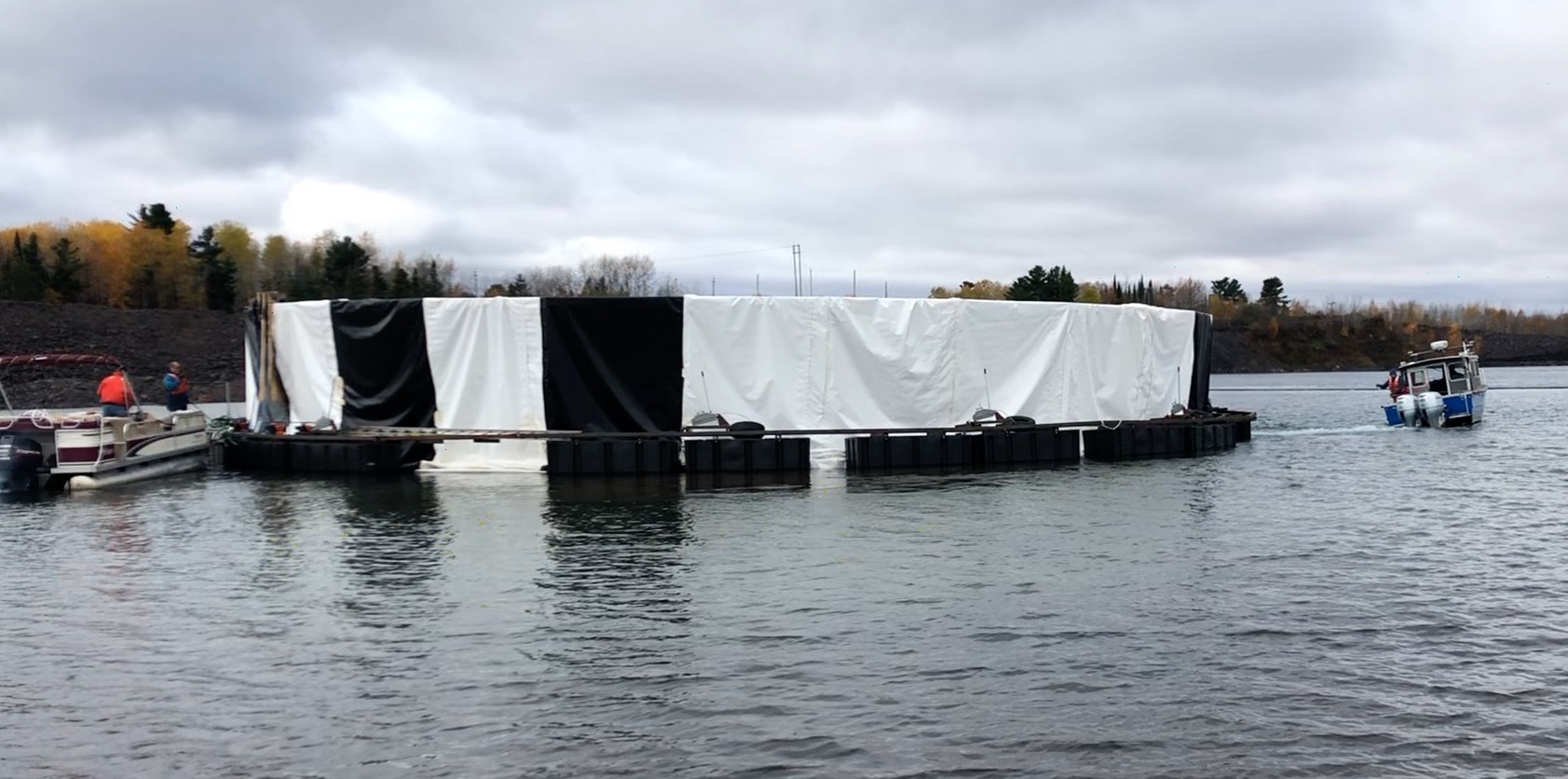}
        \caption{The detector, towed from the construction site to the deeper water.}
        \label{fig:Tow}
    \end{figure}

    The deployment procedure was planned as follows:
    \begin{enumerate}
        \item The detector would be built on dry land and sealed. %When the pit water level rose, the detector was flooded and remain in the location of its assembly.
        \item External floats surrounding the detector would be attached to support the detector during deployment
        
        \item The detector would be towed by boat to its deployment location (shown in Figure~\ref{fig:Tow}).
        \item The umbilical pipe containing the power and data connections would then be connected to the pass-through plate. The two high-density polyethylene (\textsc{hdpe}) pipes for circulating the water would also be connected so that water could be added and removed from the detector using electric pumps.
        
        \item The detector would then be filled with water from the mine pit. This would cause its bottom cap to sink while the top cap remained on the surface. Consequently, the detector height would start increasing as more water was added. This process would continue until the Dyneema cables were under tension and the detector reached its maximum height.
        \item Finally, the detector would be lowered to the bottom of the mine pit using external floating winches attached to the external float assembly. The weight of the bottom cap would keep the detector in place; the buoyancy of the top cap would keep the top cap floating above the bottom cap to give the detector its height. Overall the detector would remain sunken. At this point, the detector would be fully deployed.
    \end{enumerate}
    
    \section{Water Purity}
    \label{chap:TheCHIPSDetector:sec:WaterPurity}
    
    To maintain the water's optical clarity inside the detector, the water had to be filtered.
    
    The filtration system for the detector was not unusual. The pH of the water averaged 8 (approximately the same as sea water), there were no unusual chemicals to be filtered out and no exotic wildlife living in the mine pit. Once a year an algae bloom lasting a few days occurred on the surface of the lake but not at the bottom, therefore the filtration had to tolerate large surge quantity of dead algae. The temperature at the bottom of the lake remained a few degrees above freezing all year round.
    
\begin{figure}
        \centering
        \includegraphics[width=0.5\textwidth]{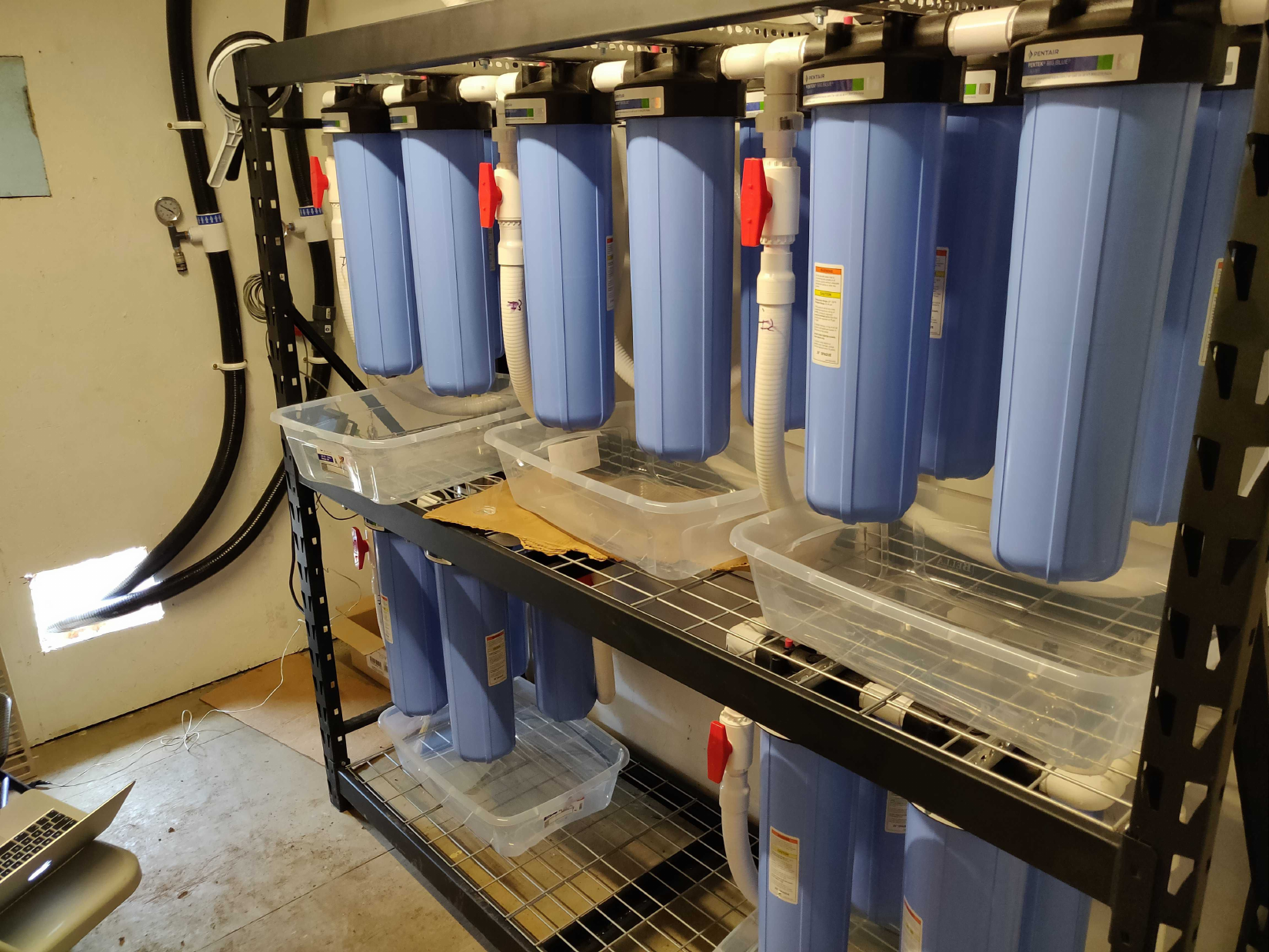}
        \caption{The interior of the water hut showing the filtration plant.}
        \label{fig:Water_Hut_Interior}
    \end{figure}

     The filtration plant was housed in a hut on the shore of the pit, the hut housed ten pairs of \SI{10}{\micro\metre}~carbon block filters and \SI{0.5}{\micro\metre}~polypropylene filters installed in parallel \cite{Campbell2020} (shown in Figure~\ref{fig:Water_Hut_Interior}).
    
    Water was to be pumped from the detector through the filters and back to the detector with a submerged pump through \textsc{hdpe} pipe. Two lengths of \textsc{hdpe} pipe were prepared by butt welding segments together until each was over~\SI{300}{\metre} long. During nominal operation, the water system formed a closed loop, so the water could be continuously cycled through the filters. To fill the detector for the first time during deployment, the water had to be pumped into the detector from the lake via the filtration plant. 
    
    Laboratory prototyping indicated that attenuation lengths in excess of~\SI{100}{\metre} were achievable with this system starting from samples of pit water \cite{Campbell2020,Amat2017}, significantly larger than the \SI{25}{\metre}~detector diameter. The system was designed to pump up to half a million litres per day; this was achieved during testing. At such rate, the full detector volume would pass through the filters approximately every two weeks. Generally, about 7~turnovers are required in order to ensure that 99\%~of the water actually passes through the filters~\cite{Battersby2022}. This would lead to complete filtering of the detector water every 3 to 4 months.
    
    \begin{figure}
        \centering
        \includegraphics[width=0.5\textwidth]{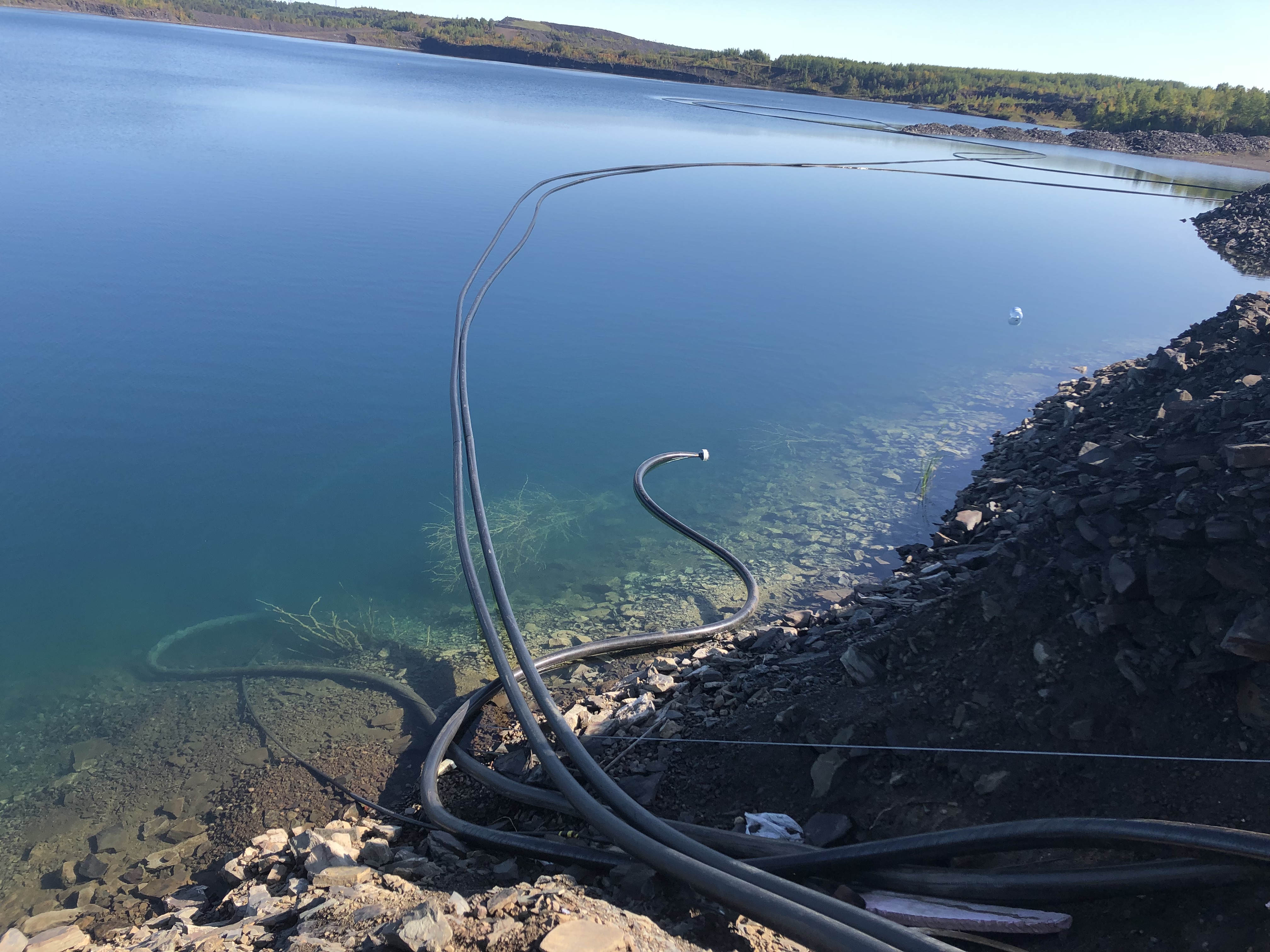}
        \caption{The completed inflow and outflow water pipes for the detector. The third, shorter pipe is the water intake from the mine pit, which was used for filling the detector during deployment.}
        \label{fig:Long_Pipe}
    \end{figure}
        
    \section{Detector Planes}
    \label{chap:TheCHIPSDetector:sec:DetectorPlanes}
    
    To observe Cherenkov light produced by neutrino interactions in water, the detector was instrumented with 2000 \textsc{pmt}s arranged into small rectangular detector planes, each carrying approximately 30~\textsc{pmt}s. These detector planes provided a modular and rigid mounting system for the \textsc{pmt}s in manageable numbers and also a waterproof connection to the readout and data acquisition (\textsc{daq}) electronics.
    
\begin{figure}
        \centering
        \includegraphics[width=\textwidth]{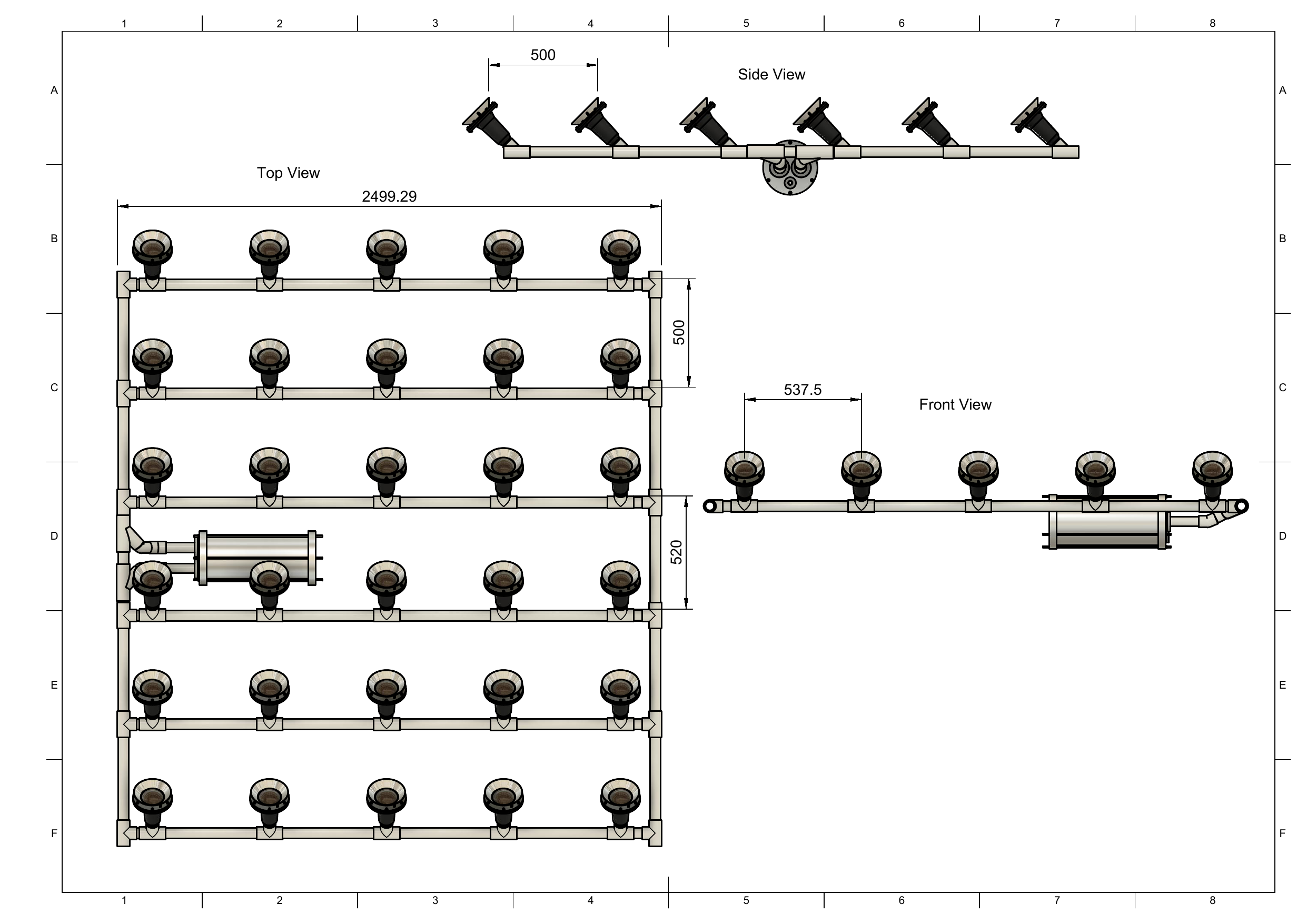}
        \caption{}
        \label{fig:plane_diagram}
    \end{figure}

    The detector planes consisted of a frame made of rungs of standard plumbing \textsc{pvc} pipe. At the centre of the frame was a cylindrical container, which housed electronics, and across the frame, pointing outwards, were plumbing fixtures, where \textsc{pmt}s were installed. Each \textsc{pmt} would be sealed inside a plastic insert that was glued to the frame, from there it was connected to readout electronics using a standard \textsc{Cat}-5~cable. This provided all of the necessary power and signalling. A rendering of a detector plane can be seen in Figure~\ref{fig:plane_diagram}.
    
    In order to detect photons, \textsc{pmt}s require high voltage (\textsc{hv}). In \textsc{Chips}, each \textsc{pmt} generated \textsc{hv} in situ using a dedicated base that implemented a Cockroft-Walton (\textsc{cw}) voltage multiplier. This meant that no large central \textsc{hv}~supply was required and no \textsc{hv}~cabling was necessary. \textsc{pmt} bases also digitised analog signals, which were distributed up-stream to each plane's \textsc{daq} electronics. The only connection between each \textsc{pmt} and the electronics container was the single \textsc{Cat}-5~cable. 
    
    Since detector planes were composed of standard \textsc{pvc}~plumbing pipes and fixtures, their assembly was greatly simplified. Waterproof seals were achieved using commercially available \textsc{pvc}~primer and cement. This was desirable in practice, as assembly could be performed on-site without specialist staff.
    
\begin{figure}
        \centering 
         \begin{subfigure}[t]{0.49\textwidth}
             \centering
             \includegraphics[width=\textwidth]{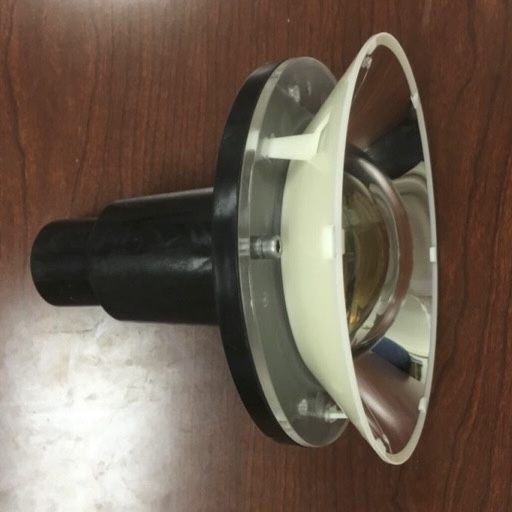}
             \caption{Assembled Nikhef \textsc{pmt}, ready to be glued onto a plane.}
             \label{fig:PMT_Assembled}
         \end{subfigure}
         \hfill
         \begin{subfigure}[t]{0.49\textwidth}
             \centering        
             {\includegraphics[width=\textwidth]{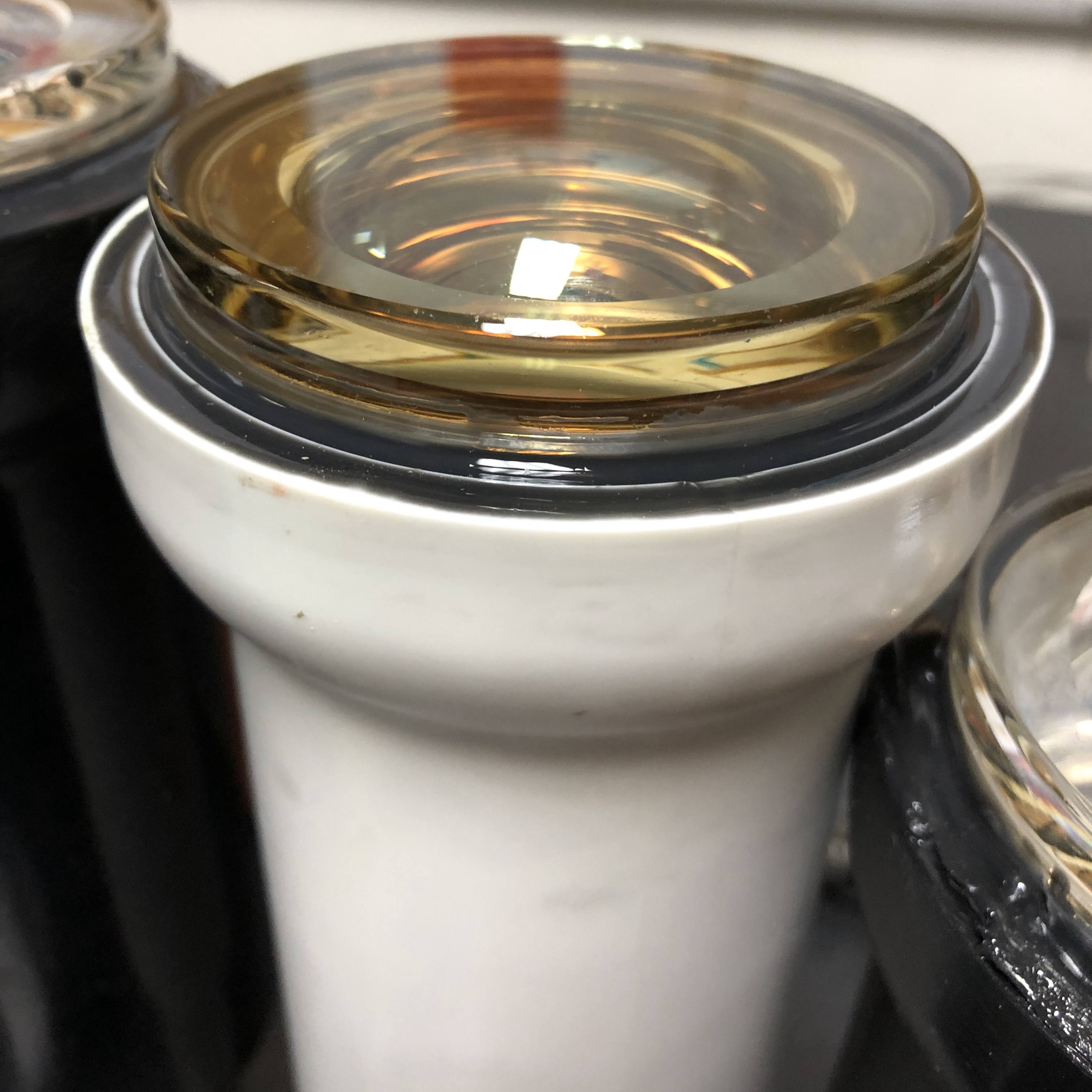}}
             \caption{The photocathode of a Madison \textsc{pmt} exposed to the environment. The tube was glued into the plastic insert (white) with a resin potting compound (black). This created a waterproof seal between the glass and the plastic.}
             \label{fig:Madison_Insert_Glue}
        \end{subfigure}
        \caption{}
        \label{fig:pmt_mounting_types}
    \end{figure}

    There were two kinds of detector planes in \chips{}, named after the locations at which they were created:
    \begin{itemize}
        \item{Nikhef planes: based on \textsc{Km}3\textsc{n}e\textsc{t} electronics \cite{Margiotta2019KM3NeTSoftware} modified for the \chips{} geometry. These planes carried \textsc{Hzc} photonics \texttt{XP82B20FNB} \textsc{pmt}s, which used a negative \textsc{hv}~supply. Consequently, during nominal operation the \textsc{pmt} face could become energised and therefore could not come into contact with water. These \textsc{pmt}s had reflectors to increase the effective detecting area. A Nikhef \textsc{pmt} can be seen in Figure~\ref{fig:PMT_Assembled}.}
        
        \item{Madison planes: designed by \chips{} at the University of Wisconsin-Madison in conjunction with the IceCube experiment \cite{Huber2017TheUpgrade}. These planes carried Hamamatsu \texttt{R6091} \textsc{pmt}s, which used positive \textsc{hv}~supply. This meant that during operation there was no potential difference between the photocathode and the water, therefore the face of the \textsc{pmt} was exposed to the water as shown in Figure \ref{fig:Madison_Insert_Glue}.}
    \end{itemize}
    
    Since the faces of the \textsc{pmt}s in the Madison planes came into contact with water, a positive \textsc{hv} was applied via a \textsc{cw} base (rather than a negative \textsc{hv}). Custom \textsc{pvc}~inserts were sealed with a watertight potting compound as shown in Figure~\ref{fig:Madison_Insert_Glue}. This allowed \textsc{pmt}s to be exposed to the detection medium with no barriers or interfaces, yet the rest of the \textsc{pmt} electronics and the plane were fully protected by a watertight seal. In contrast, \textsc{pmt}s of Nikhef planes were fully insulated from the water owing to their negative polarity \textsc{hv}. In order to minimize undesirable optical properties, acrylic was used as a transparent insulator. The \textsc{pmt}s were placed in a custom \textsc{pvc} insert and a transparent acrylic cover and a water sealing O-ring were bolted to the front. This insulated the photocathode at \textsc{hv} from the earthed water. Finally, a reflective cone was attached to the cover which increased the light collection efficiency by 60\%.
    
\begin{figure}
        \centering
         \begin{subfigure}[t]{0.49\textwidth}
             \centering
             \includegraphics[width=1.0\textwidth]{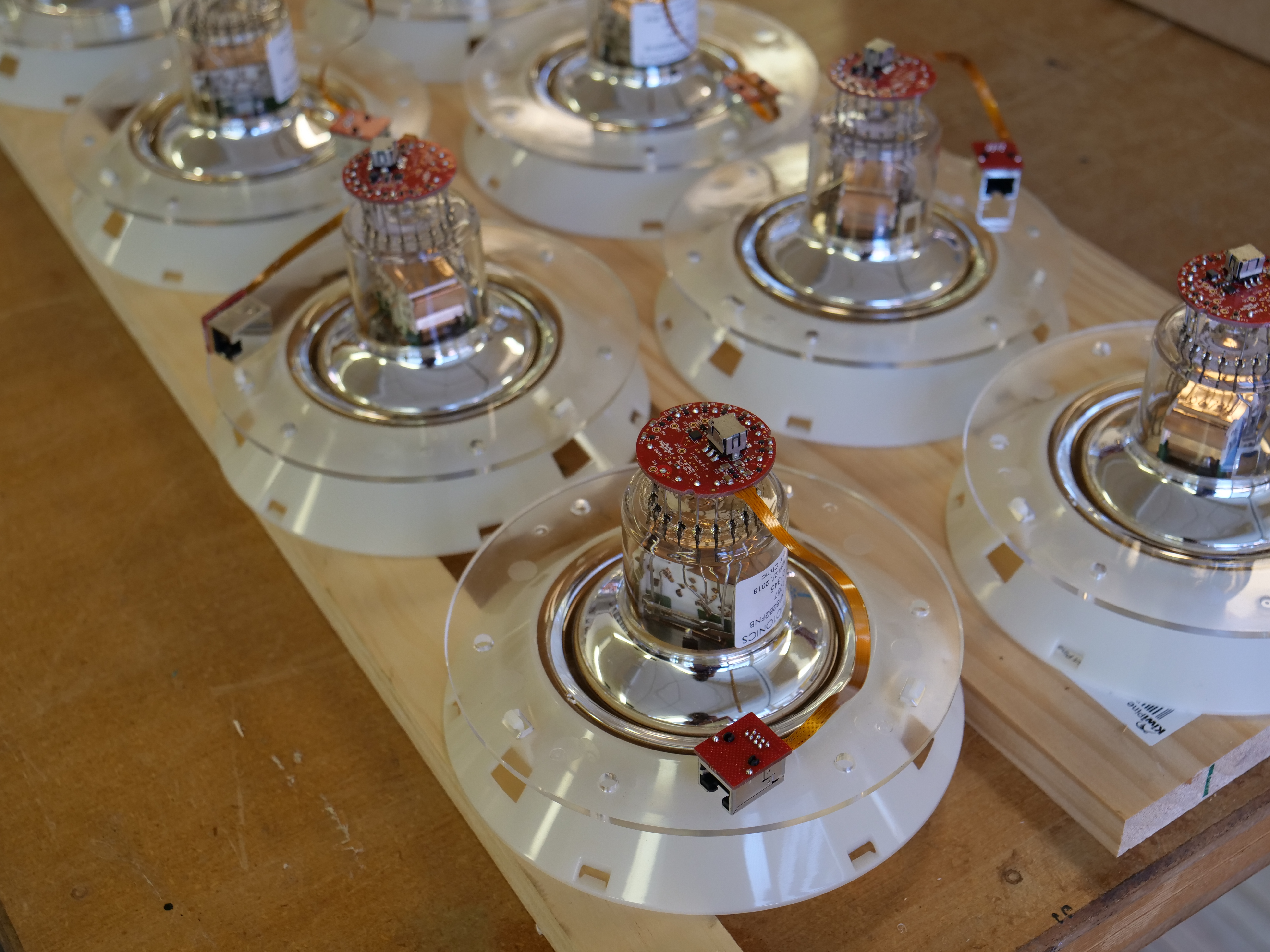}
             \caption{\textsc{pmt}s after the stabilising jig had been removed. The \textsc{pmt} covers are oriented towards the ground, so that each \textsc{pmt}s can be seated inside its cover.}
             \label{fig:Nikhef_Potted_PMT_Face_Down}
         \end{subfigure}
         \hfill
         \begin{subfigure}[t]{0.4\textwidth}
             \centering
             \includegraphics[width=1.0\textwidth]{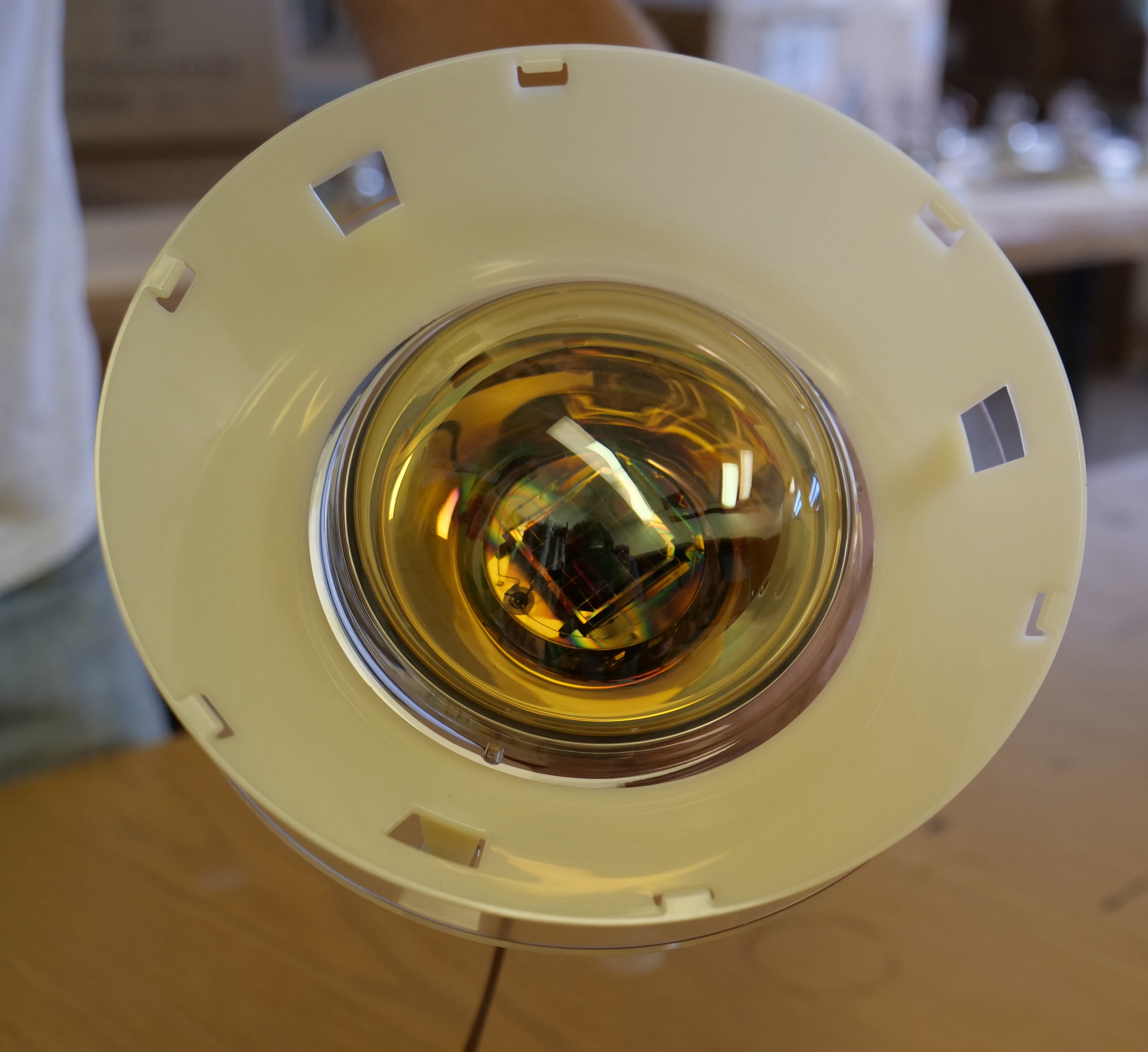}
             \caption{A potted Nikhef \textsc{pmt}. The presence of the optical glue between the plastic and the glass reduced the reflectivity between the two layers. The acrylic cover electrically insulates the face of the \textsc{pmt} from its surrounding environment.}
             \label{fig:Nikhef_Potted_PMT_Face_Up}
        \end{subfigure}
        \caption{}
        \label{fig:Potted_PMTs}
    \end{figure}

    To reduce optical losses due to reflection and refraction at the optical interface between the cover and the glass face of Nikhef \textsc{pmt}s, the glue that permanently joined these components was carefully selected to match their refractive index. It was critical that the glue was applied with no contamination, imperfections or bubbles. Furthermore, due to its relatively high cost it was desirable to minimise the amount of applied glue per \textsc{pmt}.
    The attachment of a \textsc{pmt} cover was performed as follows. First, \SI{15}{\milli\littleLitre} of potting compound was deposited inside an acrylic cover, into which a \textsc{pmt} (shown in Figure~\ref{fig:Nikhef_Potted_PMT_Face_Down}) was placed face down. A dedicated jig was developed to hold the \textsc{pmt} vertical, allowing the tube to move vertically. The natural buoyancy lifted the \textsc{pmt} slightly above the plastic, allowing the glue to occupy the space between the two media. The use of buoyancy of the \textsc{pmt} elegantly removed the need for complex supporting structures that would hold the \textsc{pmt} at a precise height, as the buoyancy of the tube was reliable and reproducible. This method produced 225 potted \textsc{pmt}s a day, which was performed by a team of students. A completed \textsc{pmt} can be seen in Figure~\ref{fig:Nikhef_Potted_PMT_Face_Up}.
    
\begin{figure}
        \centering
        \includegraphics[width=0.7\textwidth]{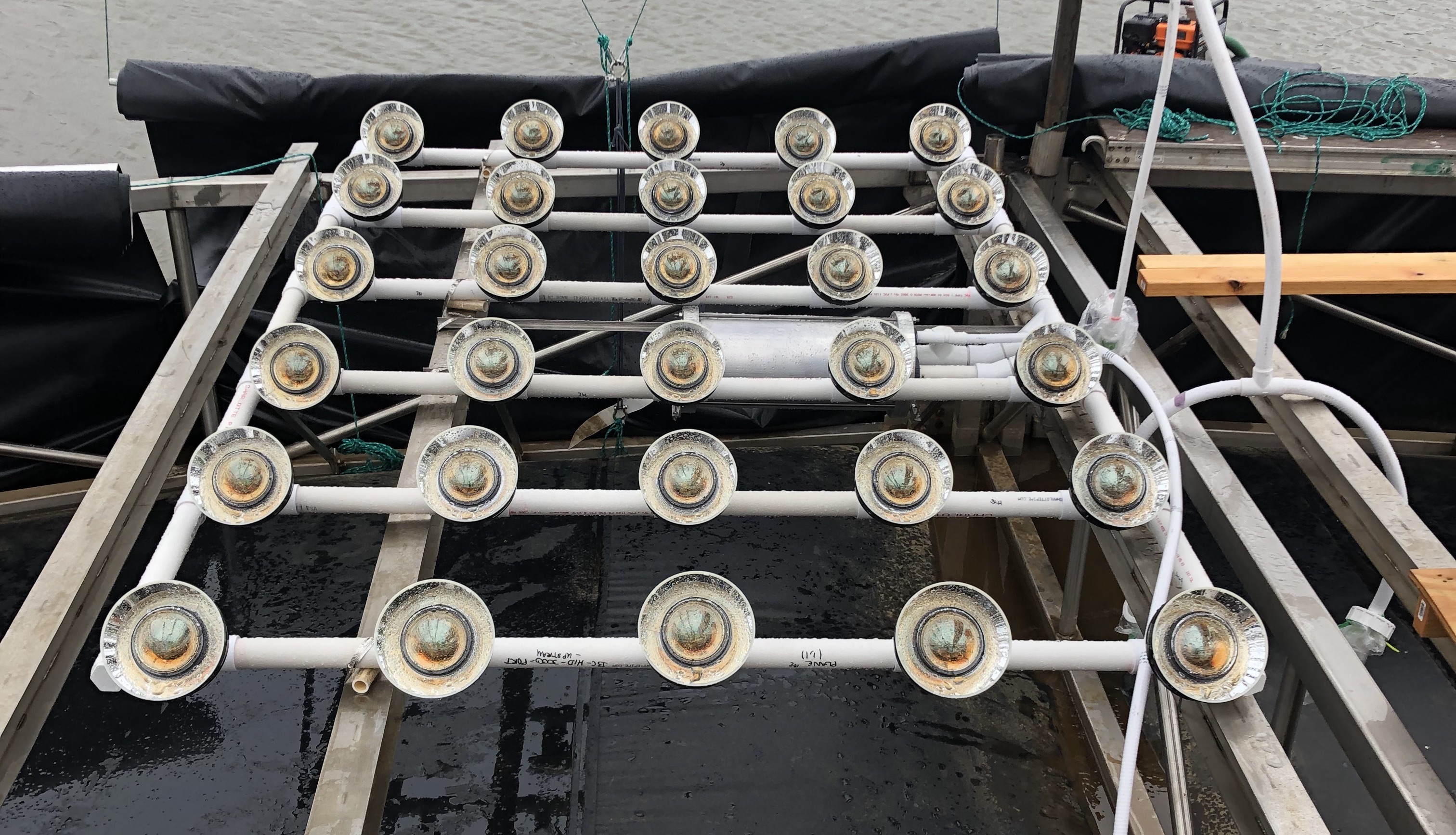}
        \caption{Completed Nikhef plane installed in the detector. Each \textsc{pmt} has an acrylic cover and reflector attached. The \textsc{pmt}s are sealed inside their inserts and installed into the \textsc{pvc} plane frame. The electronics container located in the centre of the plane. A flexible \textsc{pvc} hose (incomplete) exits the electronics container, which would connect the plane to upstream power supplies and \textsc{daq} electronics.}
        \label{fig:Plane_In_Detector}
    \end{figure}

    In total, 62~planes were placed into the detector, 56~Nikhef planes spread out between the top and bottom caps, and 6~Madison planes in just the bottom cap. Approximately 60~more Nikhef planes and 24~more Madison planes were prepared for a detector upgrade which would have increased the \textsc{pmt} coverage in the detector. An example of a completed plane installed inside the detector can be seen in Figure~\ref{fig:Plane_In_Detector}.
    
    \FloatBarrier
    
    \section{Electronics and Timing}
    \label{chap:TheCHIPSDetector:sec:ElectronicsAndTiming}
    
    Each detector plane type was operated using dedicated \textsc{daq} electronics. Madison planes used a microprocessor-based MicroDAQ board \cite{Huber2017TheUpgrade} mounted on individual \textsc{pmt}s to digitise, timestamp, record and distribute observed signals. At the centre of each plane was a fanout board which simply aggregated data streams from all MicroDAQs and multiplexed them into the detector network.
    
    Nikhef planes also used intelligence mounted on the \textsc{pmt} bases to digitise signals and fanout boards to aggregate data. However, in contrast Nikhef bases were less autonomous than MicroDAQs. This meant that a lot of functions performed by MicroDAQs were carried out by the fanout board instead (for instance, hit timestamping). These electronics were purchased from \textsc{Km3n}e\textsc{t} (with different connectors) and can be seen in references \cite{Gajanana2013ASICDetector,Real2016DigitalKM3NeT,Margiotta2019KM3NeTSoftware}.
    
    To integrate heterogeneous detector planes into a single distributed measuring system, a second level of fanout boards was introduced. These were housed in dedicated electronics containers, which were physically distributed within the detector. Each of these containers housed power and data network infrastructure implemented by White Rabbit (\textsc{wr}) switches \cite{Lipinski2011InternationalSymposium} \cite{Tomasz2011PreciseTime}. This was a \SI{1}{\giga\bit}~fibre optic Ethernet network (consisting of single-mode fibres and wavelengths of \SI{1310}{\nano\metre} and \SI{1550}{\nano\metre}), which served a dual purpose: in addition to carrying conventional data packets it transmitted precise time signals, which allowed individual sensing components of the detector to be synchronised within \SI{100}{\pico\second}. The network coalesced at a central electronics container, which was responsible for power switching and facilitating the master data link to the shore via the umbilical that was terminated at a hut on the shore. Inside the hut, a cluster of Linux computers controlled the detector, monitored its state and saved data. Also on the shore was a GPS antenna with a receiver connected to a \textsc{wr} grandmaster clock that provided absolute time reference.
    
    Implementation of the timing synchronisation and distribution system is beyond the scope of this paper. See \cite{Bash2022Low-latencyDetector}.
    
    \section{Detector Deployment and Lifecycle}
    \label{chap:TheCHIPSDetector:sec:WhatHappenedNext}
    
\begin{figure}[h]
        \centering
        \includegraphics[width=0.9\textwidth]{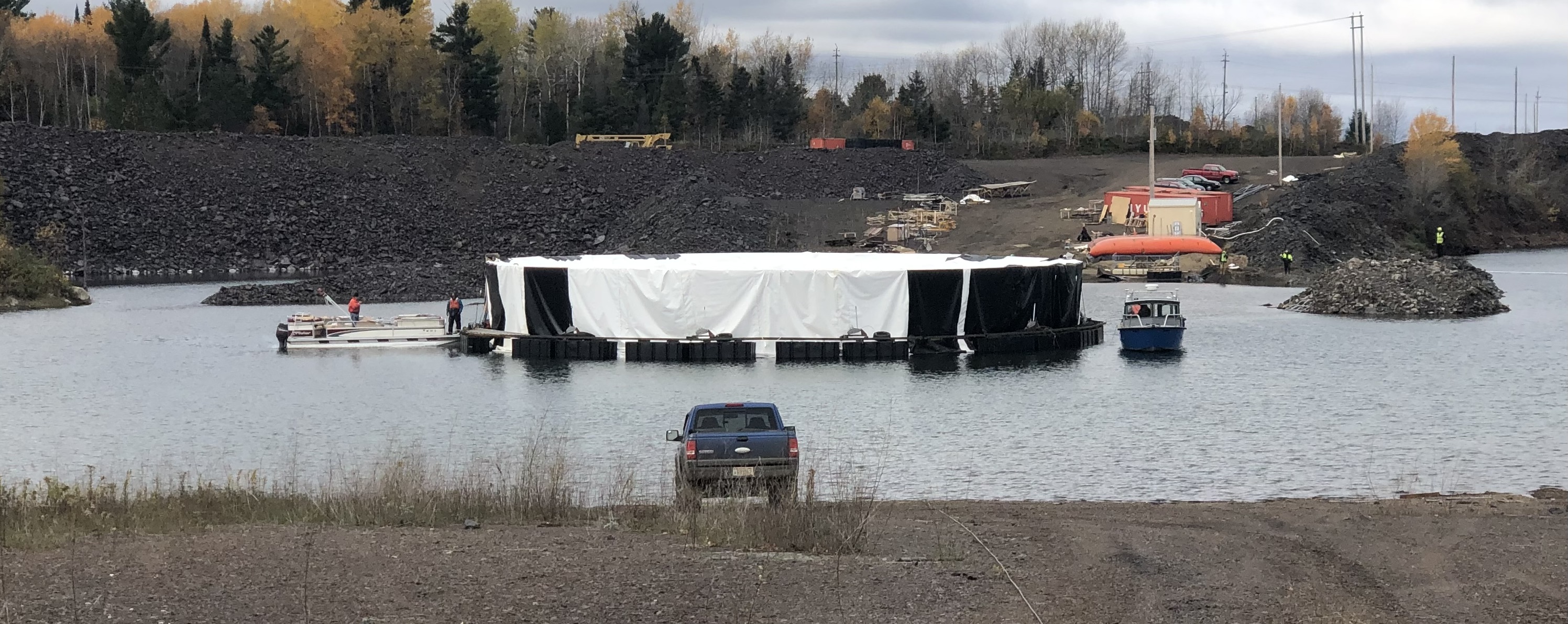}
        \caption{The exterior of the detector after completion. The liner can be seen supporting the detector and the floating dock forms a perimeter around the detector. Two boats tow and steer the detector during deployment. The scale of the detector is visible.}
        \label{fig:detectorOutsideComplete}
    \end{figure}

    The \textsc{Chips} detector was deployed in October 2019, as shown in Figure~\ref{fig:detectorOutsideComplete}. During deployment, the liner sustained damage during the towing procedure prior to water ingress. This prevented the detector interior from maintaining full buoyancy. Consequently, the float-mounted winches were supporting the full load without the advantage of additional flotation from the detector itself and partial failure of the winch system followed. The outcome was that the detector could not be fully deployed as planned before the winter. The liner was to be repaired in spring the following year when the lake thawed.
    
    \FloatBarrier
    
    In March 2020, the \textsc{Covid-19} pandemic halted all access to the site. During spring and summer, the structural condition of the system further deteriorated. Several months later, in September the detector was removed from the water as there was no clear timeline to resume activities and the pandemic had not resolved. 
    
    %also the \textsc{N}u\textsc{mi} beam shutdown was extended \cite{FermiNationalAcceleratorLaboratoryFermilabPlan}. 
    
\begin{figure}
        \centering
        \includegraphics[width=1.0\textwidth]{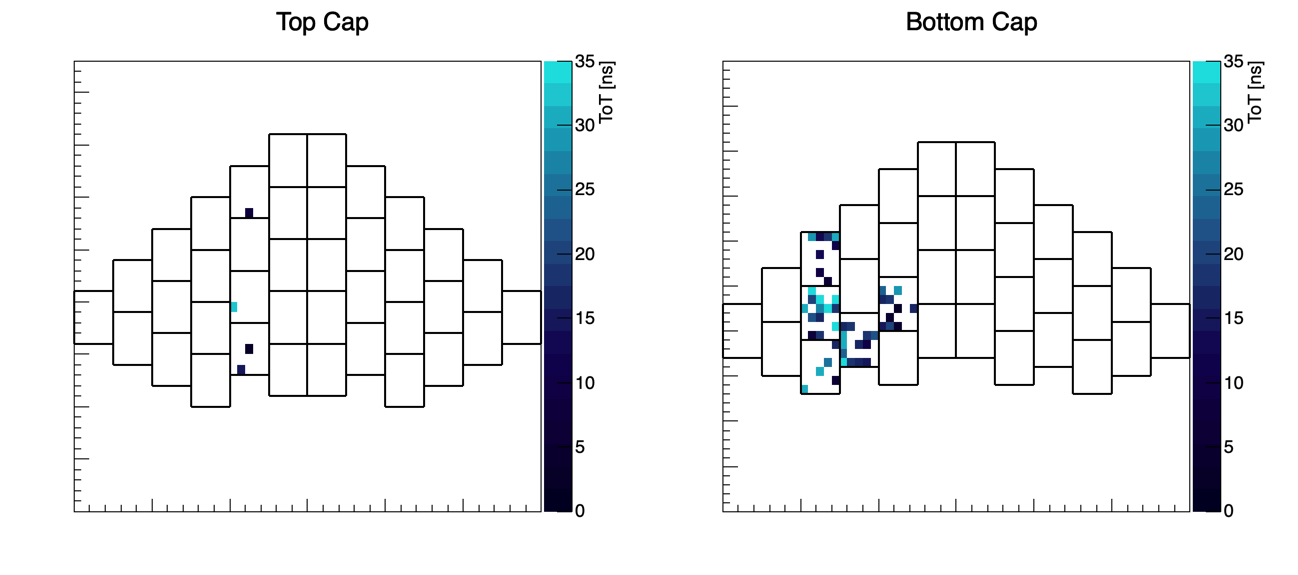}
        \caption{A \chips{} hitmap (the detector from above) showing what is likely to be a cosmic ray muon event. The damage sustained to the detector during deployment gave little time to assess detector efficiency. This event is assumed to be a cosmic ray as it is unlikely to be noise because each plane had been tested before installation to ensure low noise. At the time of this event, the \textsc{N}u\textsc{mi} beam was not running.}
        \label{fig:hitmap}
    \end{figure}

\begin{figure}
        \centering
        \includegraphics[width=0.7\textwidth]{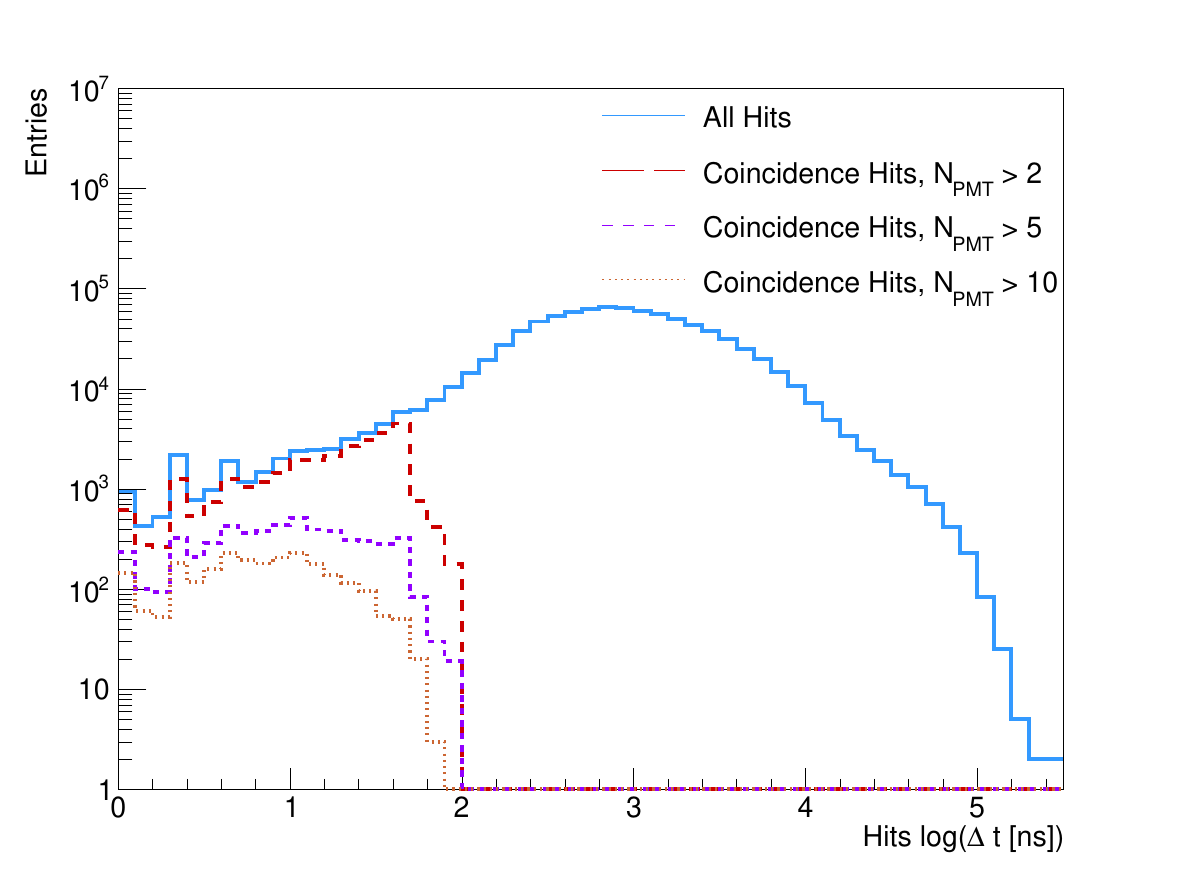}
        \caption{Time window length between two successive hits as a function of number of \textsc{pmt}s hit for different hit number thresholds. The noise is clearly seen in blue above a two hit window length of \SI{30}{\nano\second}, (note logarithmic scale) whereas the actual coincidences can be seen below that. }
        \label{fig:coincidence}
    \end{figure}

    In spite of mechanical damage that prevented its full deployment, the detector was put into limited operation for a short period of time. During that time, it underwent various technical checks that demonstrated its viability ahead of planned commissioning. For instance, the control and monitoring system, trigger and time distribution systems were successfully tested in late~2019. Furthermore, both \textsc{daq} software as well as electronics were used to take data after deployment was aborted. An example of a hitmap for a cosmic ray event can be seen in Figure~\ref{fig:hitmap} which demonstrated that the detector was operational after deployment. Figure~\ref{fig:coincidence} shows the number of \textsc{pmt}s registering hits inside a time window of \SI{30}{\nano\second} for different hit number thresholds. 
    
    Following its removal, the detector was disassembled in such a way that permitted its instrumentation to be reused for future projects. The mine site was returned to its original state.

    \section{Post-Deployment Analysis}
    \label{chap:performanceAndAnalysis}
    
    The construction of the detector was intended to take place over a Spring and Summer with the majority of the work taking place outdoors. The prior winter was longer and colder than expected with the last snow of the season arriving May 1st. Furthermore, rain, storms (structural work couldn't take place with a risk of lightning) and high temperatures made outdoor work difficult as not only the detector needed protecting but the workers did too. Overall the effects of working outdoors were underestimated and the exposed construction site was more difficult than expected. However, the construction site being outdoors and directly at the deployment location did indeed save infrastructure costs and simplify the deployment process - nevertheless future detectors would ideally be constructed under cover and shielded from the elements.
    
    The deployment process was both successful and terminal for the life of the detector. The detector was successfully built on dry land, floated, towed to the deployment location and then sunk. However, the liner was damaged during the towing process and the sinking did not go smoothly. Post-deployment analyses suggested that the detector was moved too early and not enough water had been pumped out before the towing began. The detector was still too deep so grazed against an underwater berm made of rocks which scraped a gash in the liner.The floating dock was designed to support a static vertical load in centre and not a detector still being moved sideways so was not designed to support the weight of the entire detector under such conditions. With more time, practice and preparation, a smooth deployment is certainly possible. Delays in the construction due to weather meant that the deployment time window was narrow and was done in sub-optimal conditions. Furthermore, the floating dock should have been designed for the worst case scenario of supporting the entire detector in the event of a major structural failure.
    
    The liner demonstrated both positive and negative traits. The liner was easy to weld despite the outdoor conditions and producing light liner-to-liner tight and waterproof seals was relatively easy to perform reproducibly - two previous deployments of prototypes demonstrated this. It was difficult to the make the light tight seals attaching the liner to the steel frame. It was learned that bolting the liner to steel and sealing the liner with rubberised gaskets leaked both light and water. Sealing the liner between two steel 
    plates while also using rubber washers produced better seals but reliability studies were not performed in detail due to time constraints and the low number of steel-to-liner attachment points. The rip in the liner during deployment was unfortunate but given the enormous forces involved scraping the detector against rock a tear was to be expected. The liner was strong and the material was fit for purpose but greater care should have been taken to protect it and this was a major limitation of the experimental setup. Future experiments should consider the same material but protect it accordingly.
    
    The steel frame was a major success of the experiment. Despite being heavy, the steel frame was strong enough that the detector could be supported asymmetrically and remain completely rigid and certainly saved the detector during deployment. Furthermore, the frame was rigid enough that each end cap could be lifted out of the water by crane with just four tie-in points without buckling which greatly simplified the decommissioning of the experiment.
    
    \section{Conclusion}
    \label{sec:conclusion}
    
    The \chips{} R\&D project aimed to demonstrate that significant cost reduction could be achieved in construction of \textsc{wc} neutrino detectors. To this end, a prototype was designed and built using widely available and inexpensive components. Its assembly and deployment were carried out in a disused mine pit with a workforce that, on the whole, was non-specialist. During project realisation, a variety of cost-saving methods were tested, refined and shown to be viable for large-scale application. Consequently, the total cost of the hardware components for the prototype was estimated to be €1.7m which could have equipped \SI{15}{\kilo\tonne} if the detector had been expanded fully. Despite not observing accelerator neutrinos during its relatively short lifespan, the detector's instrumentation passed numerous internal tests, indicating its suitability for future scientific operation and demonstrating the capability to observe directed light in a large-scale \textsc{wc} detector. The total cost of the hardware is estimated to be on approximately 
    
    In summary, the construction of the \chips{} prototype entailed broad exploration of the landscape of cost reduction strategies, which could be potentially employed beyond the scope of this project in future low-cost neutrino detectors. In that sense, even though the prototype did not observe \textsc{N}u\textsc{mi} neutrinos, vast majority of goals set for this project were satisfied at the time of its deployment. The prototype detector was assembled and deployed from April to October 2019. The detector was decommissioned in July 2020. Its instrumentation is currently reused for other research projects.

    \section{Acknowledgements}
    \label{sec:acknowledgements}
    
    This work was supported by Fermilab; the Leverhulme Trust Research Project Grant; U.S.~Department of Energy; The Royal Society Research Professorship funding for J.Thomas and the European Research Council funding for the CHROMIUM project.
    
    The \textsc{Chips} collaboration would like to thank the Wisconsin IceCube Particle Astrophysics Center, University of Wisconsin–Madison, University College London, University of Alberta, University of Minnesota Duluth, Aix-Marseille University. Additionally, the \textsc{Chips} collaboration would like to thank PolyMet and Cleveland-Cliffs for hosting the experiment.
    
    Fermilab is operated by Fermi Research Alliance, LLC, under Contract No. DE-AC02-07CH11359 with the U.S.~DOE.

    \pagebreak
 
\bibliographystyle{elsarticle-num}

\bibliography{references.bib}

\end{document}